\documentclass[aps,pra,showpacs,showkeys,onecolumn,notitlepage,groupedaddress]{revtex4-2}

\usepackage{graphicx,tabularx,times,hyperref}

\setcitestyle{numbers,square}
\usepackage{array}
\usepackage{amsmath}
\usepackage{mathrsfs}
\usepackage{units}
\usepackage{graphicx}
\usepackage{braket}
\usepackage{xfrac}
\usepackage{enumitem}
\usepackage[normalem]{ulem}
\usepackage{booktabs}
\usepackage{tabularx}
\usepackage{tikz}
\usepackage{multirow, bigstrut}
\usepackage{hhline}
\usepackage{times}
\usetikzlibrary{shapes.geometric, arrows, matrix, positioning}
\usetikzlibrary{fit}
\tikzset{%
highlight/.style={rectangle,rounded corners,draw,
fill opacity=0.5,thick,inner sep=0pt}
}

\usepackage{hyperref}
\usepackage{pdfpages}

\usepackage{etoolbox} 

\makeatletter
\patchcmd{\@outputpage@head}{\@ifx{\LS@rot\@undefined}{}{\LS@rot}}{}{}{}
\makeatother

\hypersetup{
colorlinks=true,
linkcolor=cyan,
filecolor=magenta, 
urlcolor=blue,
}

\tikzstyle{startstop} = [rectangle, rounded corners, minimum width=3cm, minimum height=1cm, text centered, text width=5cm, draw=black]
\tikzstyle{io} = [rectangle, rounded corners, minimum width=5cm, minimum height=2cm, text centered, text width=5cm, draw=black]
\tikzstyle{goal} = [rectangle, rounded corners, minimum width=5cm, minimum height=2cm, text centered, text width=9cm, draw=black]
\tikzstyle{iog} = [rectangle, rounded corners, minimum width=5cm, minimum height=2cm, text centered, text width=5cm, draw=black, fill=gray!30]
\tikzstyle{goalg} = [rectangle, rounded corners, minimum width=5cm, minimum height=2cm, text centered, text width=9cm, draw=black, fill=gray!30]
\tikzstyle{arrow} = [thick,->,>=stealth]

\begin{document}

\title{Exploring pedagogical content knowledge of physics teaching assistants using the Energy and Momentum Conceptual Survey}

\author{Liam Doyle*}
\affiliation{Department of Physics and Astronomy, University of Pittsburgh, Pittsburgh, PA 15260}
\author{Alexandru Maries}
\affiliation{Department of Physics, University of Cincinnati, 
Cincinnati, OH 45221, USA}
\author{Chandralekha Singh}
\affiliation{Department of Physics and Astronomy, University of Pittsburgh, Pittsburgh, PA 15260, USA}
\thanks{Corresponding author: lcd23@pitt.edu}

\begin{abstract}

This study examines the extent to which physics graduate teaching assistants (TAs) are aware of introductory physics student thinking and the types of challenges introductory students commonly have with energy and momentum concepts, which is important for implementing active learning methodologies and supporting diverse learners in physics courses. We present findings from a TA professional development course and discuss an approach to investigate TAs' pedagogical content knowledge, specifically their ability to recognize introductory student conceptual difficulties.
We investigated 70 first-year graduate TAs' ability to identify common introductory physics student alternate conceptions on the Energy and Momentum Conceptual Survey (EMCS). The TAs participated in a professional development course that emphasized reflection on introductory student thinking patterns to promote evidence-based pedagogical practices. TAs predicted the most common incorrect answers introductory students would select after lecture-based instruction, then compared their predictions with actual data from introductory students followed by a class discussion. This reflection process can promote robust understanding of introductory student thinking patterns, which is essential for effective teaching that accounts for how introductory students make sense of physics concepts. Results reveal gaps between TAs' perceptions and introductory student thinking, with TAs performing poorly on many of the analyzed questions. For example, TAs consistently overestimated that introductory physics students would use novice-like thinking in many situations posed in EMCS problems, i.e., they expected introductory physics students to make more novice-like errors than they actually did. These findings have important implications for effective instructional design, e.g., physics TAs who misunderstand introductory physics student capabilities may inadvertently create barriers to learning by spending valuable class time on either over-scaffolding or under-challenging introductory physics learners and not spending time on pedagogical issues that are important to address.
\end{abstract}
\maketitle

\section{Introduction and Framework} \label{Intro}

Higher education faces unprecedented challenges in preparing educators who can facilitate effective learning experiences. There are many issues to address, for example, difficulties in attracting and retaining teachers \cite{bell2019synthesis, van2020teacher, wilson2022landscape, love2023teacher}. This difficulty is in part due to a lack of effective professional development programs that can improve teachers' perceptions of their own teaching competence, which can lead to improved retention \cite{song2021stem, AAAS}. 
In addition, students are increasingly enrolling in college courses with more diverse backgrounds, resulting in an increased emphasis on using integrated approaches to science, technology, engineering, and mathematics (STEM) education \cite{ryu2019preservice, lo2021design, margot2019teachers, morrison2021teachers, berisha2021developing, lane2022bridging, hamad2022understanding, christian2021ngss, kelley2020increasing}, further highlighting the importance of professional development \cite{goldhaber2019evidence, bartels2019shaping}. 
Unfortunately, STEM courses are often taught using traditional lecture-based instruction, which fails to address the varied ways learners in these courses construct knowledge; thus, proper preparation of educators on the facilitation of effective learning practices is vital. This can be seen in studies that discuss professional development courses that provide educators with experiences that help them implement an integrated approach to STEM education \cite{kurup2019building, wu2019scaffolding, mumcu2023teacher, brown2019professional, dare2021beyond, boice2021supporting}, how professional development helps educators \cite{wright2019developing, weinberg2021professional, bardelli2023teacher, akiri2022professional}, and investigations on professional development courses and how to improve them \cite{goldhaber2022front, karisan2019effect, chai2019teacher}.

Graduate teaching assistants (TAs), who provide an important component of instruction in undergraduate STEM courses, represent a critical leverage point for educational transformation. Despite this, most TAs receive minimal pedagogical preparation for their teaching responsibilities even though such preparation can lead to significant benefits to student learning \cite{shum2021learner,lee2019impact,Huffmyer01012019}. 
It has been found that TAs tend to associate less with being teachers and more with being tutors or lab managers \cite{zotos2020investigation}, while other studies note the benefits when TAs have positive perceptions of teaching \cite{wheeler2019exploratory, goodwin2021enthusiastic}.
Smith and \textit{et al.} \cite{smith2023exploring} found that TAs with higher self-efficacy are better with student-centered approaches than those with lower self-efficacy. Therefore, professional development classes, which teach about pedagogical preparation and help to develop TA self-efficacy, similar to the ones provided to other educators, could have positive impacts on TAs as they begin their teaching journeys as well as the students who rely on their instruction.

While TAs are an important part of the instruction in many universities, they are a group that is studied less than educators at the K-12 and university level. 
Therefore, this study focuses on physics graduate TAs and emphasizes how systematic development of their pedagogical content knowledge (PCK), specifically their understanding of introductory physics student thinking patterns, discussed here in the context of energy and momentum, provides one type of foundation for them to be able to implement effective teaching practices. As this work is a part of a group of projects focusing on the pedagogical content knowledge of teaching assistants \cite{tugkPCK, fciPCK, csemPCK}, the theoretical framework and methodology for all of these projects are similar. As a result, the  beginning sections bear similarities to the three previous research studies \cite{tugkPCK, fciPCK, csemPCK}. 

\subsection{Background on Graduate Teaching Assistants}

Graduate students play a vital role in educating the next generation of students. At universities around the United States, it is typical for physics graduate TAs to teach introductory recitations or lab sections, which have smaller enrollments than the lecture portion of the course (typically, these sections have approximately 20--40 students, compared to the 100+ in the lecture portion of the course). Recitations supplement lectures and typically include a short review of relevant topics covered earlier in lecture, provide students with the opportunity to practice problem solving and ask questions on content covered (including on homework problems) that they would otherwise be unable to due to the often large enrollment sections in lecture. Occasionally, recitations may include a quiz to encourage student participation.

Thus, TAs are valuable members of a teaching team involved in large classes because they have the unique position of being more easily accessible to students. This unique position suggests that it is important to provide professional development to TAs to assist them in performing their duties in a manner conducive to student learning. There is a wealth of research on TAs investigating different aspects of teaching: TAs' beliefs and practices in teaching \cite{singh2009categorization, marshman2017grading, tasolutionbeliefs,signchange,kajfez, sandi-urena,broken,conflicting,dannyta,Good_2024}, the importance of receiving buy-in from TAs about using evidence-based instructional methods \cite{bauer, goertzen, wilcox}, descriptions and evaluations of professional development programs for training TAs \cite{otero2010physicsdeptrole,lawrenztrainingTA, sandifer2015recruiting, armenti, bozak, Etkina2000, Flaherty, marshmantypesofproblems, gretton, linenberger, marbach-ad, miller2014, roehrig, Sharpe2000}, TAs learning about collaborative learning \cite{ghimire1,ghimire2}, and others \cite{Chini&AlRawi2012, Ethington, FeldonScience, french, wyse}. Reviewing these studies in detail is beyond the scope of this article. For a short review of the literature on effective TA programs, see, e.g., Ref. \cite{MariesTAs2020}.

\subsection{Pedagogical Content Knowledge Framework} \label{PCK framework}

Pedagogical content knowledge \cite{shulman1986PCK, shulman1987PCK} refers to subject matter knowledge for teaching and has been used often by researchers in education \cite{van1998developing, grossman1990makingteacher, grossman1991we, gess2001examining, loughran2004search, borko1995expanding, ebert1993assessment, geddis1993transforming, van2002development, zavala2007innovative, zollman1994preparing, akkocc2010investigating, carter1993place, kagan1990goldilocks, loughran2012understanding, baxter1999assessment, yang2014study}. Shulman \cite{shulman1986PCK, shulman1987PCK} defined PCK as a type of knowledge that instructors use when making practical decisions about how to teach a lesson on a specific topic. In Shulman's view, PCK incorporates knowledge about the kinds of representations of concepts and analogies that are conducive to learning, illustrations and examples, and an ``understanding of what makes learning of specific topics easy or difficult" \cite{shulman1986PCK}. In other words, an important aspect of PCK is knowledge about common student difficulties because having a sense of where students struggle is essential to helping them develop a robust knowledge structure. 
In particular, educators should know where to focus instruction and what to pay attention to within their instructional design in order to help students overcome common pitfalls. 
Awareness of common student difficulties can also prove invaluable when designing appropriate scaffolding to help students develop facility with the concepts and their multiple representations. For example, previous research with K-12 teachers \cite{sadler2013middleschool} found that on questions with strong distractors, there is a large difference in learning gains between students who were taught by teachers who could successfully identify the alternate conceptions and students taught by teachers who could not. Thus, it is important to explore the extent to which teachers are knowledgeable about common student difficulties.

There have been many attempts by researchers to document teachers' PCK and how it develops \cite{loughran2004search,loughran2000science,ebert1993assessment, geddis1993transforming, van2002development, akkocc2010investigating}, though the majority of this research is in the context of K-12. However, these tasks have proven difficult because teachers' knowledge of their practice is often tacit \cite{carter1993place, kagan1990goldilocks}; that is, a few observations are often not sufficient in providing a complete picture of all the tools at teachers' disposal when deciding how to approach teaching a specific topic. This means that extended observations are needed to recognize when the teachers' PCK is utilized during instruction \cite{loughran2004search}, as well as interviews to understand what PCK was not used and why. Multi-method approaches have been used to overcome these challenges to understand what goes into developing the coursework. In particular, the array of instructional decisions has been investigated, such as what representations to use, what examples and types of activities students engage in, etc. For example, Loughran  \textit{et al.} \cite{loughran2004search} used classroom observations and follow-up interviews with teachers. These interviews encouraged teachers to further articulate their knowledge alongside alternate representations that were not necessarily used during instruction. This investigative approach is time consuming to carry out and analyze since there are lots of qualitative data that come out of the observations and interviews, all of which require coding and analysis. For a review of methods used to study teachers' PCK, see Ref. \cite{baxter1999assessment}.

Though many studies have been done investigating the PCK of K-12 teachers, graduate TAs may have different levels of both subject knowledge and PCK compared to K-12 teachers. We note that very few studies have focused on the PCK of graduate student TAs, specifically, their knowledge of common student difficulties that we investigate here, despite their importance in many of these large enrollment courses for universities. One example comes from biology, where Lampley et al. \cite{Lampley2018PCKBio} discuss a professional development program for biology TAs via lesson study. Lesson study refers to a system of investigation that helps teachers refine their ideas about best teaching practices through classroom-level research. In the initial phase, it includes discussions of common misconceptions that teachers expect to occur for a specific topic being taught, as well as a plan for data collection and analysis to determine the extent to which the instruction was successful and to inform future instructional changes. The TAs in the study worked in groups to design, teach, and improve their lessons. They found this type of professional development helped improve TAs' beliefs about biology instruction as well as certain aspects of their PCK, including knowledge of specific common misconceptions.

Park and Rizzolo \cite{park2025PCKmath} investigated mathematics TAs' expectations of student difficulties. They provided TAs with students' written work and asked them to identify where students were having difficulty, create a plan to address the difficulties they identified, and implement the plan in the classroom. Their study found that the TAs often identified procedural difficulties rather than conceptual difficulties, and this informed the approaches they used in teaching. Specifically, they tended to favor a ``teaching by telling approach" rather than using the difficulties as resources for learning, suggesting that in order to help TAs develop into effective teachers, professional development should provide guidance for TAs so that they are able to understand students' conceptual difficulties (rather than procedural), as well as guidance on effective ways to incorporate common difficulties in instruction.

In physics, Thompson et al. \cite{thompson2011preparing} describe a professional development program that incorporated readings from educational literature focused on student ideas. The students in the program were prospective teachers, some physics graduate students, and others from a nonphysics background, meaning that many of the students in the program did not have the physics content knowledge, so the course also taught them the content by making use of instructional strategies advocated by the physics education research community. Part of the assessment of the course included activities in which the prospective teachers had to ``generate [incorrect] hypothetical student responses to unfamiliar questions," sometimes more than one per question. In other words, the prospective teachers were provided with a physics question and asked to provide an incorrect student response based on their understanding of student ideas that they had developed from the course readings on the literature of student difficulties. It is noteworthy that they also collected data on the prospective teachers' knowledge of student ideas before instruction, though the questions used were significantly easier, and the authors describe changes in the prospective teachers' knowledge of student ideas from before to after instruction. However, this study provided few details about the prospective teachers' knowledge of student ideas (both before and after instruction) and focused on providing evidence that there was significant improvement with near-ceiling performance on the post-test (for questions on electric circuits). The authors also note that this investigation was in the preliminary stage, but to our knowledge they have not followed up with another article that provides more details.

Even though the three articles discussed above do not provide specific details on the graduate TAs' pedagogical content knowledge, they can be very useful for those interested in developing a professional development program that explicitly incorporates approaches to help develop TAs' PCK regarding common student difficulties. It is our hope that the method we discuss here, along with the data presented in this article for both introductory students and graduate TAs, can be helpful in designing activities to support TAs' PCK of student difficulties.

\subsection{Prior Research in Other Content Areas Employing the Approach used Here}

We developed a relatively straightforward method for delving into one aspect of PCK, namely, educators' knowledge of common student difficulties with particular topics. This method, which has been utilized in three prior studies \cite{tugkPCK, fciPCK, csemPCK}, uses standardized conceptual multiple-choice surveys developed by physics education researchers and quantitative data from introductory students who took those surveys. TAs are provided with a copy of a particular survey (e.g., the Force Concept Inventory \cite{fciPCK}), and for each question on the survey, they are asked to select what they expect to be the most common incorrect answer choice selected by introductory physics students after having received traditional lecture-based instruction on the topic. Then, quantitative student data are used to quantify the extent to which the TAs are knowledgeable about common student difficulties. 
Three prior research studies conducted with TAs using the method described above used the Test of Understanding Graphs in Kinematics (TUG-K) \cite{tugkPCK}, the Force Concept Inventory (FCI) \cite{fciPCK}, and the Conceptual Survey of Electricity and Magnetism (CSEM) \cite{csemPCK}. This method is also currently being applied in two other studies involving the Survey of Thermodynamic Processes and First and Second Laws \cite{stpfaslPCK} and the Rotational and Rolling Motion Conceptual Survey \cite{rrmcsPCK}. The main findings of these studies are as follows:

\begin{itemize}
    \item TAs' knowledge of students' alternate conceptions is context dependent; that is, they can readily identify a particular alternate conception in certain contexts, but not in others. 
    \item TAs sometimes expect certain alternate conceptions to be very common when they are in fact not at all common among introductory students.
    \item Think-aloud interviews with TAs attempting to determine students' alternate conceptions suggested that the TAs were reflective and had reasonable thoughts with respect to potential reasoning students may use.
    \item The TAs struggled more in identifying alternate conceptions in the context of the CSEM than in the context of the FCI or TUG-K. This may be due to the fact that many alternate conceptions in the context of mechanics come from people's experience with the real world and thus may be easier to predict than alternate conceptions with electricity and magnetism, where the concepts are more abstract.
\end{itemize}

As noted in Sec. \ref{PCK framework}, there is very little research on the PCK of TAs, specifically, on their understanding of common student difficulties with physics content, and the study presented here, along with the ones summarized above, aims to fill that gap.

\subsection{Goal of this Study}

Based upon prior research findings, it is very clear that it is valuable to investigate physics TAs' knowledge about common introductory student difficulties in concrete contexts to learn about situations in which the TAs have a good understanding of introductory physics student difficulties and those in which they struggle to identify common student difficulties. As Shulman and others have argued \cite{shulman1986PCK, shulman1987PCK, Berry2016}, knowledge of common difficulties in different content areas is one part of PCK, and PCK serves as the primary guide when teachers decide how to implement instruction and what to focus on. 
To improve TAs' PCK about common student difficulties, a professional development course could make use of an activity similar to what we describe here in which the TAs attempt to determine the most common student difficulties on multiple-choice questions taken from validated conceptual surveys. After working on this task on their own, there could be a small-group or whole-class discussion in which the TAs discuss each question and how they think introductory students may answer it which includes what incorrect reasoning students can use. Furthermore, the TAs could be shown student data, which provides information about which incorrect options were most common (e.g., 40\% of students select an incorrect option), and the TAs could discuss what type of student reasoning may lead students to select that option. Such an activity can be quite valuable in enhancing TAs' knowledge of student difficulties.

These considerations, along with the valuable findings of previous research \cite{tugkPCK, fciPCK, csemPCK}, motivated us to carry out the research study discussed here focusing on the PCK of physics TAs related to energy and momentum concepts using the Energy and Momentum Conceptual Survey (EMCS) \cite{EMCS,brundage,rebelloemcs}. The EMCS  is an assessment instrument designed to evaluate students' conceptual understanding of introductory energy and momentum concepts. Thus, the goal of this study was to evaluate one aspect of TAs' PCK, namely, their understanding of common introductory physics student alternate conceptions of energy and momentum concepts included in the EMCS.

\section{Methodology} \label{Methods}

The methodology for this study discussed below is similar to our 
previous studies which were focused on other conceptual physics topics using conceptual surveys \cite{tugkPCK, fciPCK, csemPCK} (and those that occurred concurrently to this study \cite{stpfaslPCK, rrmcsPCK}).

\subsection{Participants and Context of the Study}

The participants in this study were 70 first-year graduate students (from four separate cohorts) enrolled in a semester-long mandatory pedagogy-oriented TA professional development course at a large research university. These cohorts ranged from as low as 13 to as high as 25 graduate students. These graduate students come from many different backgrounds, and approximately 30\% of them are women. While roughly 50\% of the graduate students in these cohorts completed their undergraduate education in the United States, very few of them (less than 10\%) did so at the institution in which this study was conducted. We note that we did not separate these cohorts, as a previous study had already compared these groups and found that there was no significant difference between those who had and had not completed their undergraduate education in the United States \cite{tugkPCK}.

Since all the graduate students were in their first year, they were completing the typical core graduate physics courses (e.g., Quantum Mechanics, Classical Mechanics, Mathematical Methods, and Statistical Physics). Typically, for the first two semesters, these graduate students take 3--5 courses per semester. They are required to take four core courses (usually two per semester) in order to continue to Ph.D. candidacy. While these first-year graduate students are able to engage in research, it is usually recommended that they focus on coursework in their first year, so it is very rare for them to be working in a research lab, though the majority work as TAs (most commonly in labs or recitations).

The TAs in this course typically teach recitations and labs, usually in a traditional manner. It is important to note that the TAs who teach recitations generally have significant freedom regarding how to carry out their duties, though this is dependent on the instructor they are working with. Regardless, many TAs' recitations tend to follow a broad guideline of answering student questions on homework, solving practice problems with students, and allowing for questions and discussion as they do so, followed by a quiz. Depending on the instructor, the quiz may be taken individually or in small groups, but other than these general outlines, the TAs tend to have flexibility with how they hold recitation. This suggests that TAs who are knowledgeable about effective instructional approaches can have a significant impact on student learning. In the labs, the TAs begin by demonstrating the procedures needed for the lab, and then the students closely follow the detailed procedures written in the lab manual with guidance from TAs when they have difficulties. There are also a few graduate students in this course who are not yet assigned a TA position, but they will have a TA position at least for two semesters at some point in their graduate education.

The professional development course meets once a week for two hours and is the only pedagogy-oriented course that many of these graduate students will take, so it is designed to introduce TAs to many different aspects of teaching in order to help them become effective teachers. At the beginning of the course, the graduate students learn about cognitive research and physics education research  and consider how to implement what they learn from this research in a classroom environment. Following this, each week, the TAs complete different reflective exercises designed to help them perform their TA responsibilities in a student-centered manner. For example, in one class, the TAs are asked to write a homework problem and have a classmate work out the problem from a student perspective and provide feedback. Other classes focus on other activities such as writing solutions to problems that will be provided to students and discussing what features solutions should have to help students learn, grading student work in a manner that is conducive to promoting use of effective problem-solving strategies, designing problems for recitation/homework/quizzes/exams, and discussing the considerations leading to different types of scaffolding that may be used in different contexts. In the second half of the semester, each TA is assigned to create an example problem on a particular topic and present the problem solution to the other graduate students who are acting as undergraduate students. They are encouraged to ask questions that may come from students who are learning this topic for the first time. This gives each TA the opportunity to practice leading a discussion in recitation and field questions from students about a topic that's new to them. The TAs are then able to get feedback on their teaching and their process of answering questions. Furthermore, part of the goal of the course is to help TAs recognize the importance of being knowledgeable about student difficulties so that they can assist these students in developing expertise, so typically, there are multiple classes in which the TAs attempt to identify common student difficulties on specific conceptual surveys, the EMCS being one of them. These classes always include showing the TAs quantitative data from introductory students and discussions that focus on specific student difficulties and how they could be addressed during instruction.

\subsection{Materials}

The materials used in this study are the EMCS itself (which can be found Supplemental Materials), the post-instruction data from 457 calculus-based introductory physics students across four separate equivalent courses, and TA data about their PCK using the method that will be described in Sec. \ref{Methods}, and the qualitative interview data obtained from the TAs. The post-instruction data from calculus-based physics students was collected by asking students to complete the EMCS and identify the correct answers. These data were used to determine students' alternate conceptions on each question of the EMCS, therefore allowing the assessment of the extent to which the TAs are able to identify these alternate conceptions (described in detail in Sec. \ref{Methods}).

We note that the EMCS uses language that aligns with the language used in typical introductory physics courses and textbooks, e.g., that the mechanical energy of the system is conserved when there is no work done on the system by non-conservative forces. Some educators have advocated instead using language that the mechanical energy of a system is ``constant" instead of ``conserved," since mechanical energy is not a conserved quantity, but rather one which is constant in certain situations. Also, some have suggested refraining from describing some forces as ``non-conservative" since all forces are conservative at a fundamental level \cite{SeeleyExamining2019, SeeleyUpdatingLanguage2022, EtkinaDesigning2018, ChabayUnifiedApproach2019}. Therefore, the language used in this paper will align with the recommended language rather than using the same language as the EMCS.

\subsection{Methods}

The TAs in the first two cohorts were provided a copy of the EMCS, and for each item on the EMCS, they did the following:
\begin{itemize}
    \item identified the correct answer,
    \item identified which incorrect answer choice they expected would be the most commonly selected by introductory students in a post-test (i.e., after traditional lecture-based instruction in relevant concepts),
    \item provided a brief explanation of their reason for selecting a given option.
\end{itemize}
We refer to this as the ``PCK task."

While giving similar PCK tasks that used different conceptual surveys \cite{stpfaslPCK, rrmcsPCK} to the first two cohorts, it was observed that for some of the questions, some of the TAs were unable to figure out the correct answer. Therefore, it was decided that after identifying the correct answers on their own, the TAs should be provided with the correct answers. Specifically, for the last two cohorts, the PCK task was slightly different. For each item on the EMCS, the TAs in the last two cohorts:
\begin{itemize}
    \item identified the correct answer,
    \item were provided the correct answer,
    \item identified which incorrect answer choice they expected would be the most commonly selected by introductory students in a post-test (i.e., after traditional lecture-based instruction in relevant concepts),
    \item provided a brief explanation of their reason for selecting a given option.
\end{itemize}

This change did not affect the data much, with the exception of two items, namely Q16 and Q2,3 where a sizable percentage of the TAs in the first two cohorts selected the correct answer. In the first two cohorts, roughly one quarter of the TAs on Q16 and one fifth of the TAs on Q23 identified the correct answer as the most common incorrect answer choice that introductory students would select. This is because these TAs did not know the correct answer to these questions, and this issue was fixed in the last two cohorts.

Consistent with the previous three studies \cite{fciPCK,csemPCK,tugkPCK}, the TAs were asked to identify students' alternate conceptions after instruction because it was considered that it is more important for TAs to be aware of the alternate conceptions that are unlikely to be eliminated after traditional lecture-based instruction. We note that it does not make much qualitative difference whether the PCK task is framed for before instruction or after instruction because students' alternate conceptions are mainly the same, just that after instruction, a lower percentage of students have these alternate conceptions.

In order to obtain a quantitative measure of TAs' performance in identifying the persisting alternate conceptions that remain after traditional instruction, scores were assigned to each TA for each question on the EMCS. Each incorrect answer was assigned a score that was equal to the percentage of students who selected that incorrect answer choice. These percentages came from the introductory students who were asked to identify the correct answers of the EMCS. We note that, in order to get a broader understanding of the difficulties that students have, the TAs were compared to all of the students whose data were collected for this study, not just the student data that are associated with the cohort they specifically taught.

If a TA selected the correct answer for a specific question (which was rare and only occurred for the first two cohorts, as they were not specifically provided the correct answers while completing the PCK task), they were given a normalized score of zero (calculation will be described below) because they were asked to identify the most common \textit{incorrect} answer choice of introductory students. For example, on Q1, the percentages of students who selected the incorrect answer choices A, C, D, and E, were 6\%, 34\%, 8\%, and 6\% (as shown in Figure \ref{fig:EMCS_PCK_Responses}), respectively (choice B is correct). Therefore, the scores assigned to TAs for selecting each answer choice are 0.06, 0.34, 0.08, and 0.06, respectively. (We note that we refer to these scores as ``PCK scores" because they relate to TAs' PCK about student difficulties.)

This weighting was chosen because the more prevalent a difficulty is for students, the more important it is for TAs to be aware of that difficulty so they can account for it during instruction. This weighting scheme allows TAs who identify a slightly less prevalent difficulty to still receive a good PCK score for a given question rather than giving them a low score for not identifying the most common difficulty. For example, on Q17, 25\% of students selected A and 27\% of students selected D (both incorrect); therefore a TA selected A, and their PCK score would be 0.25, and if they selected D it would be 0.27 --- two very similar PCK scores.

For each question, we calculated the average PCK score of the TAs by averaging the individual PCK scores for all TAs for that particular question (this is called ``TA avg. PCK" in Figure \ref{fig:EMCS_PCK_Responses}). However, since the maximum and minimum possible PCK scores are question dependent (e.g., on Q1, the minimum is 0.06 and the maximum is 0.34; on Q6, the minimum is 0.01 and the maximum is 0.14), we created a normalization in order to be able to easily compare TAs' performance on different questions (we refer to this as ``Norm. TA avg. PCK" in Figure \ref{fig:EMCS_PCK_Responses}). The normalization is such that it varies from 0\% to 100\%, where 0\% corresponds to all the TAs selecting the least common incorrect answer choice of introductory students (or the correct answer choice) and 100\% corresponds to all the TAs selecting the most common incorrect answer choice. The normalization for each question on the EMCS was done as follows (ave, max, and min stand for average, maximum, and minimum PCK scores on that question, respectively):

\medskip

\begin{center}
    
$    Normalized\ TA\ Ave\ PCK \ Score = 
    (\frac{ave\ question\ PCK\ score - min\ possible\ score}{max\ possible\ score - min\ possible\ score})*100\%$
\end{center}

\medskip

To give a specific example: On Q1, the scores for A, C, D, and E are 0.057, 0.337, 0.079, and 0.064, respectively (note that choice B is the correct answer and that these values have been rounded to the nearest percentage point in Fig. \ref{fig:EMCS_PCK_Responses}). The minimum possible PCK score is 0.057, the maximum possible PCK score is 0.337, and TAs' average PCK score is 0.167. Therefore, the normalized PCK score on Q1 is $(0.167-0.057)/(0.337-0.057)*100\% = 39\%$.

In short, the normalization is such that the TAs would obtain a normalized average PCK score of 0\% if they all unanimously identified the incorrect answer choice that was selected by the smallest percentage of students, and they would obtain a normalized average PCK score of 100\% if they all unanimously identified the incorrect answer choice that was selected by the largest percentage of students. (We note that as one can expect, there are no questions in which the TAs are unanimous, so there are no instances in which these extremes occur.)

To get a sense of the variation in TAs' ability to identify common introductory student difficulties on the entire EMCS survey, we first calculated each TA's PCK score on the entire EMCS survey and normalized it on a scale from 0\% to 100\%, where 0\% and 100\% would correspond to a TA who always selected the least common/most common incorrect answer choice for each question, respectively. This was done by calculating a normalized PCK score for each TA for each question and then averaging across all questions to obtain an average PCK score for each TA for the entire EMCS. The normalized PCK score for a particular question for each TA was similar to what was described earlier, except for only one TA instead of all of them (question-specific PCK score refers to the PCK score a TA received on that question, i.e., the percentage of introductory students who selected the incorrect answer choice that was identified by the TA as the most common on that question):

\begin{center}
$    Normalized \ TA\ PCK \ Score = 
    (\frac{question\ specific\ PCK\ score - min\ possible\ score}{max\ possible\ score - min\ possible\ score})*100\%$
\end{center}

We note that if a TA selected the correct answer for a particular question (a rare occurrence), they were assigned a normalized PCK score of 0\% on that question.

Lastly, to obtain further accounts of TAs' reasoning, think-aloud interviews were conducted with 6 TAs. These 6 TAs are from three of the four cohorts included in the qualitative data and had completed the in-class PCK task anywhere from three years to six months before volunteering to do these interviews. Therefore, the questions on the EMCS were not fresh in their minds when interviewed, and the interview data were used exclusively to gain a more in-depth understanding of the reasoning TAs use when engaged in the PCK task. Regarding the representativeness of these interviewees to the rest of the TA population that completed the quantitative data, we note that these interviewees were not offered financial incentives to complete these interviews. In particular, these interviewees volunteered on their own to complete this task, so it may be possible that these specific TAs are interested in teaching. These interviews were completed in approximately one hour over Zoom. During these interviews, TAs first identified the correct answer to a given question on the EMCS, then identified which incorrect answer choice they thought would be most commonly selected by an introductory student, and finally provided an explanation of why they thought that answer choice would be common. During this process, the TAs were allowed to complete this series of tasks on each question without interruption (similar to the first cohort), meaning that they were not informed of the correct answer until after they finished the full task for a given question. In other words, if a TA did not identify the correct answer (a rare occurrence), they were allowed to finish the task before being informed of the correct answer, and then they were asked if knowing the correct answer changed their selection of the most common incorrect answer of introductory students. In almost all cases where a TAs failed to identify the correct answer for a question, the TA said that the answer they previously thought was correct would likely be the most common incorrect answer choice of introductory students because the students may share the same difficulty they had, which led them to think a particular incorrect answer was correct. 

After completing the six interviews, we found that the TA responses from the interviews were very similar to those of the quantitative data. We found them to be extremely helpful for confirming the written responses as well as providing extra context to why TAs identified specific incorrect answer choices. Therefore, all the quotations in this paper are from written reasoning provided by the TAs during the in-class PCK task since interview findings were consistent with written reasoning from TAs.

\subsection{Research Questions}

In this study, we investigated the following research questions:

RQ1: How well do TAs predict introductory physics students' alternate conceptions on the EMCS after traditional lecture-based instruction?

This research question can be divided into several parts that are interconnected:
\begin{itemize}
    \item RQ1-A: What alternate conceptions that were common among introductory students did the TAs struggle to identify?
    \item RQ1-B: What alternate conceptions that were common among introductory students were the TAs able to identify?
    \item RQ1-C: To what extent are TAs able to identify alternate conceptions across different contexts (i.e., across EMCS questions)?
\end{itemize}

Our second research question pertains to comparing the results of this study with our previous studies that used similar methodologies to assess TAs' PCK, but using different physics conceptual surveys:

RQ2: What broader similarities and differences are there in TAs' ability to identify alternate conceptions on the EMCS with their ability to identify alternate conceptions on other surveys (FCI, TUG-K, CSEM) \cite{tugkPCK, fciPCK, csemPCK}?

\section{Results} \label{Results}
We note that in this section, we often refer to the ``performance of TAs", or alternatively, ``PCK score," ``PCK performance." Since our goal is to investigate the extent to which the TAs are aware of common student difficulties, ``performance" or ``score" always refers to the PCK scores calculated as described in the methods section and never to their performance in identifying the correct answers. (Also, since the TAs are all physics graduate students, they almost always identify the correct answers on the EMCS questions.)

\subsection{Performance of TAs on the Entire PCK Task}

\begin{figure}
    \centering
    \includegraphics[width=0.95\linewidth]{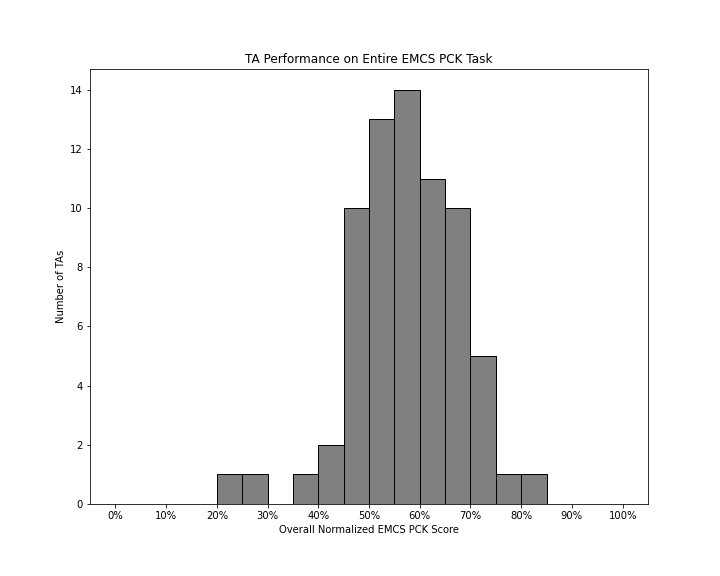}
    \caption{Performance of TAs for the entire survey. We note that the average normalized score of all TAs across the full EMCS survey was 58\%.}
    \label{fig:EMCS_TA_Hist}
\end{figure}

The distribution of average PCK scores for all the TAs is shown in Figure \ref{fig:EMCS_TA_Hist}. As seen in Figure \ref{fig:EMCS_TA_Hist}, the performance of the TAs is close to a normal distribution, where the scores of the vast majority of the TAs are bunched around the average (58\%). In particular, 90\% of the TAs had average PCK scores between 45\% and 75\%.

\subsection{Performance of TAs on the PCK Task Across the EMCS Themes}

\begin{figure}[h!]
    \centering
    \includegraphics[scale = 0.75]{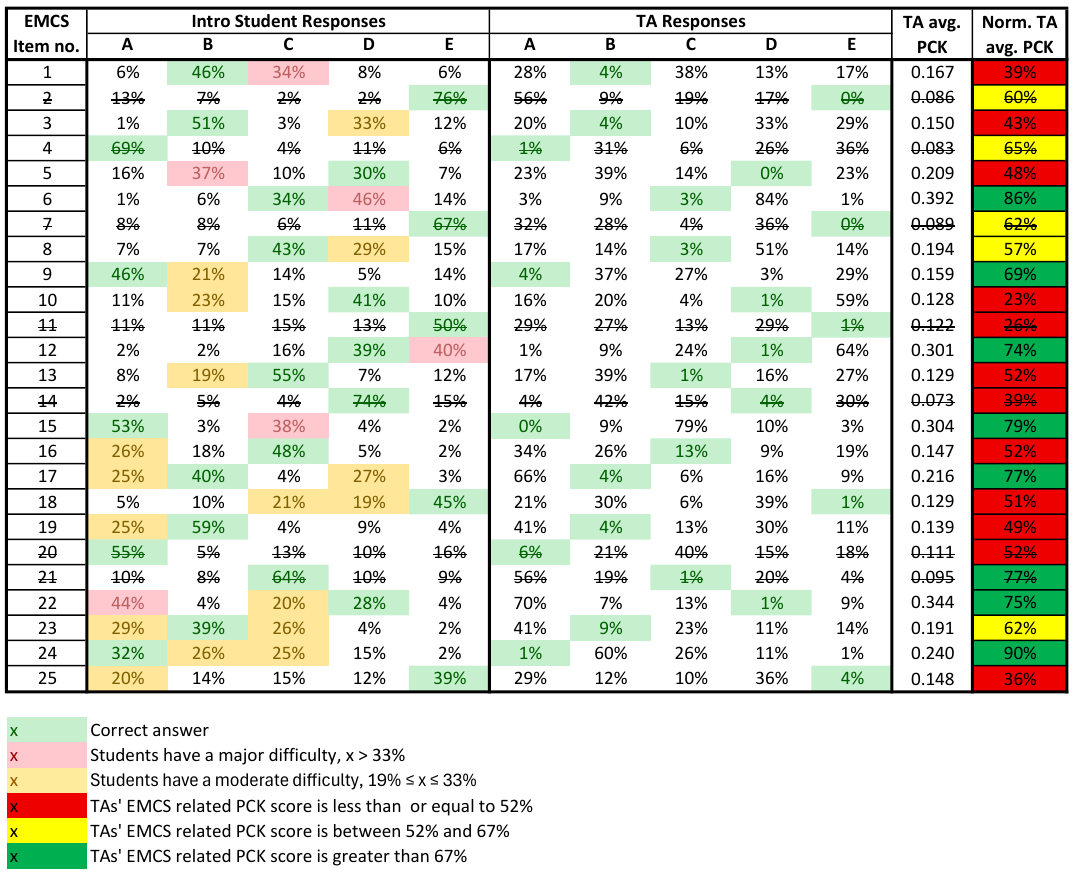}
    \caption{Questions on the EMCS with responses from students and TAs. Percentages of introductory physics students (intro student responses) who selected choices A through E in a post-test (who were asked to select the correct answer for each question after instruction). Percentages of graduate students (TA responses) who selected choices A through E (who were asked to select the most common incorrect answer that an introductory student would choose). The first column lists the question number. The second to last column lists the average graduate student PCK score (TA avg. PCK score), and the last column lists the graduate students' normalized average PCK score on a scale of 0 to 100 (Norm. TA avg. PCK). Any question that has no introductory physics students' difficulty, which reached the threshold of 19\% or more, has been left in for clarity of data because educators may still be interested in those questions, but has been struck through for ease of identification and the fact that we do not discuss them further in this paper.}
    \label{fig:EMCS_PCK_Responses}
\end{figure}

Analysis of the PCK performance of the graduate TAs was performed on each of the questions in the EMCS, which revealed common alternate conceptions as shown in Figure \ref{fig:EMCS_PCK_Responses} using the methods described in Sec. \ref{Methods}. Consistent with a similar previous study using the TUG-K \cite{tugkPCK}, we considered that the introductory students had a major difficulty if more than 33\% of them selected a particular incorrect answer choice, and that a moderate difficulty related to an incorrect answer choice selected by more than 20\% of students. 
However, there were two questions in which 19\% of students selected an incorrect answer choice, and we decided to discuss those questions as well. In the end, there were 18 questions on the EMCS with either moderate or major difficulties: Q1, Q3, Q5, Q6, Q8--Q10, Q12, Q13, Q15--Q19, and Q22--Q25. We note that all the quantitative data collected is provided in Figure \ref{fig:EMCS_PCK_Responses}. However, any question that did not reach the threshold as noted above has been crossed out so that it is easily identifiable as a question that is not discussed because students did not exhibit prevalent difficulties on that question (e.g., Q2 which was answered correctly by 76\% of students).

We note that the EMCS is provided Supplemental Materials. However, when discussing specific questions on which the context of a problem is difficult to describe without a figure, the question figure is provided near the text discussing that question.

TAs' performance was somewhat bimodal on the 18 questions with major or moderate difficulties, see Figure \ref{fig:EMCS_PCK_Responses}. In particular, on nine questions, TA PCK performance was poor, on two questions, TA PCK performance was moderate, and on seven questions, TA PCK performance was good. In other words, TAs' performance was either poor or good, with only two questions on which it was somewhere in the middle.

For the remainder of this section, we group questions based on the concepts involved in identifying the correct answer. In prior studies \cite{fciPCK,tugkPCK}, it was considered that a normalized PCK score of 50\% or less is poor. This is due to the fact that a normalized average PCK score of below 50\% indicates that, on average, the difficulties identified by TAs account for less than half of the difficulties that students have on a given question. However, when ranking questions based on normalized PCK score, there were three questions on which TAs' performance was very close to 50\% (Q18--51\%, Q16--52\%, and Q13--52\%), and the next lowest performance was on Q8 at 57\%. Therefore, we shifted the cutoff for poor performance slightly to 52\%. Also, consistent with prior studies \cite{fciPCK,tugkPCK}, TAs' normalized PCK score was considered good if it was greater than two-thirds or 67\%. Similarly, this is due to the fact that a normalized average PCK score of above 67\% indicates that, on average, the difficulties identified by TAs account for more than two-thirds of the difficulties that students have on a given question. Finally, TAs' normalized average scores between 52\% and 67\% were considered moderate. We note that when we discuss different questions on the EMCS below and we mention ``TAs' PCK score," we are referring to the normalized average PCK score to make it easy to compare different questions, and all quotations from TAs are from their written responses provided during the completion of the in-class PCK task.

\subsubsection{Mechanical Energy is Constant and Momentum is Conserved (Q3, Q16)}

Q3 and Q16 are related to elastic (Q3) and inelastic (Q16) collisions and require students to recognize when momentum is conserved and/or total mechanical energy is constant. On both of these questions, TAs' PCK score is poor --- 43\% and 52\%, as shown in Figure \ref{fig:EMCS_PCK_Responses}.  (We note that Q16 is extremely challenging, even for graduate students. In one of the cohorts in which we first asked graduate students for the correct answer for all questions before asking them to complete the PCK task, roughly one-third provided the correct response for Q16.) 

In Q3, a white hockey puck collides elastically with a stationary red hockey puck on a frictionless surface. The question asks to identify which of the following statements are true: (1) The kinetic energy of the white puck is constant (same before and after the collision), (2) the linear momentum of the white puck is conserved, (3) the linear momentum of the two-puck system is conserved. There was a major difficulty in which 33\% of students selected that (1) and (3) are the only true statements (choice D). In other words, one third of the students do not realize that it's the kinetic energy of the system of two pucks that is constant, not just a part of the system, despite realizing this regarding total momentum. Many TAs struggled to identify this alternate conception (only 33\% of them did), resulting in a low PCK score of 43\%. 29\% of TAs selected choice E as the most common incorrect answer (all three statements are true). One TA stated, ``Students may not distinguish conservation in a system [from conservation] in a body of a system."
Despite this, only 12\% of students selected this option. This indicates that the TAs who identified this incorrect answer choice thought that students may be confused about conservation of momentum because they expected students to select both statements (2) that the momentum of the white puck is conserved and (3) that the momentum of the two-puck system is conserved. This suggests that the TAs expected that students are more confused about conservation of momentum than they are because very few students selected both statements (2) and (3).
Additionally, 20\% of TAs selected choice A, which states that only statement (1) is true, but only 1\% of students selected this answer choice. One TA stated, ``As the collision is elastic, most of the students will only put emphasis on kinetic energy conservation and think choice A is correct." One could interpret that the TAs who selected A expect students to exhibit more novice-like thinking than they actually do because choice A does not include conservation of linear momentum. In other words, by selecting A as the most common incorrect answer choice of introductory students, the TAs are implicitly saying that they think that students may not realize that in a collision, momentum is conserved. However, 99\% of students selected an answer choice that mentioned momentum conservation.

\begin{figure}[h]
    \centering
    \includegraphics[scale = 0.75]{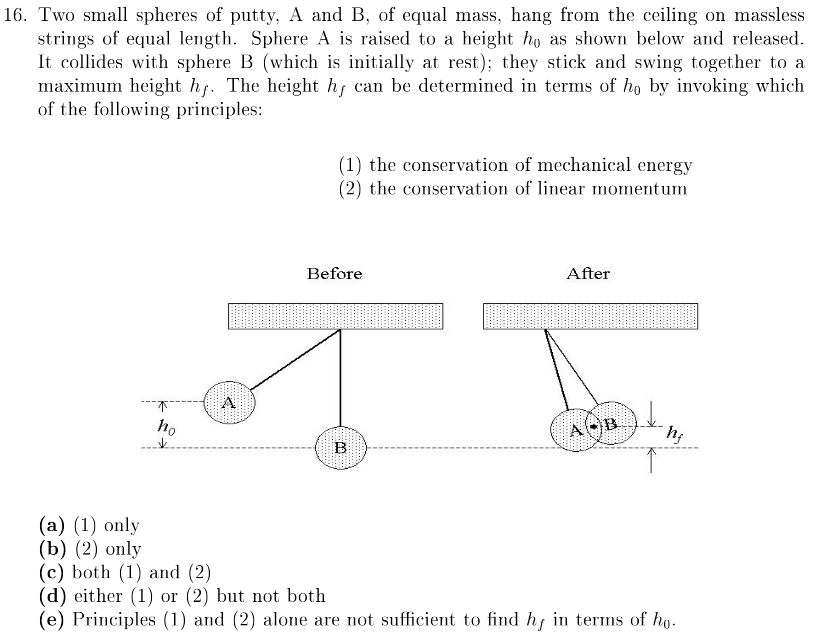}
    \caption{Figure provided for Q16 on the EMCS with the options provided for reference.}
    \label{fig:Q16}
\end{figure}

Q16 provides the diagram shown in Figure \ref{fig:Q16} and asks which principles are required to solve for the final height in terms of the initial height.
On this question, students who did not get the answer correct either mostly selected that only mechanical energy is constant is needed (26\% --- answer choice A) or that only conservation of momentum is needed (18\% --- answer choice B). The majority of the TAs (60\%) identified one of these two answer choices as the most common incorrect answer, but 19\% of the TAs selected answer choice E, which states that these two principles are not sufficient, an answer choice selected by only 2\% of students. One TA motivated their choice of E by stating, ``People think that you need to know what the initial height is to know what the final height is," and another stated, ``Students might think that the masses should be given." This suggests that the TAs who selected this answer choice thought that students may not consider the physics principles relevant to solving this problem and think that actual numbers are needed to solve the problem, which is more novice-like than they actually were. Furthermore, we should note that this question is also a bit of an outlier because 13\% of the TAs selected the correct answer as the most common incorrect answer of introductory students. This suggests that they had the same alternate conception as students (only one principle or the other is needed, not both). This was the only question on the EMCS in which more than 10\% of the TAs selected the correct answer choice, and it contributes to their low PCK performance.

\subsubsection{Conservation of Momentum (Q5, Q10)}

Q5 and Q10 both require use of conservation of momentum. TAs performed poorly on both of these questions, receiving PCK scores of 48\% and 23\%, respectively, as shown in Figure \ref{fig:EMCS_PCK_Responses}.

\begin{figure}[h]
    \centering
    \includegraphics[scale = 0.75]{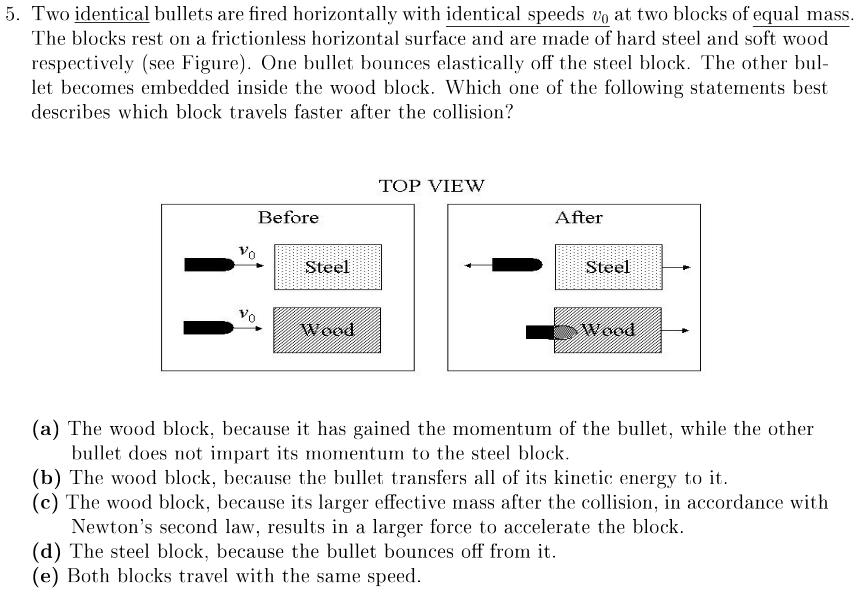}
    \caption{Q5 from the EMCS with the options provided for reference.}
    \label{fig:Q5}
\end{figure}

On Q5, two identical bullets are shot toward two blocks of equal mass (one made of steel, the other made of wood). For the wooden block, the bullet embeds in it, and for the steel block, the bullet bounces off as shown in Figure \ref{fig:Q5}. Q5 asks which block travels faster after the collision. There was a major difficulty on this question, as 37\% of students selected that the block that has the bullet embedded would travel faster because the bullet transfers all of its kinetic energy to it (choice B). While 39\% of the TAs identified this alternate conception, 23\% of them selected option E, which states that both blocks will travel with the same speed (choice E) --- an option selected by only 7\% of students. One TA who selected choice E stated, ``I believe that students will think that because the two bullets are identical, traveling with identical speeds at two blocks of equal mass then the momentum is conserved and confuse this law with their speeds also being equal." This implies that TAs who selected choice E thought students might not recognize that there is a difference between an elastic and an inelastic collision. Similar to the reasoning they used in the previous two questions discussed, it appears that on Q5 as well, the TAs may expect introductory physics students to be more novice-like than they actually are.

\begin{figure}
    \centering
    \includegraphics[scale = 0.75]{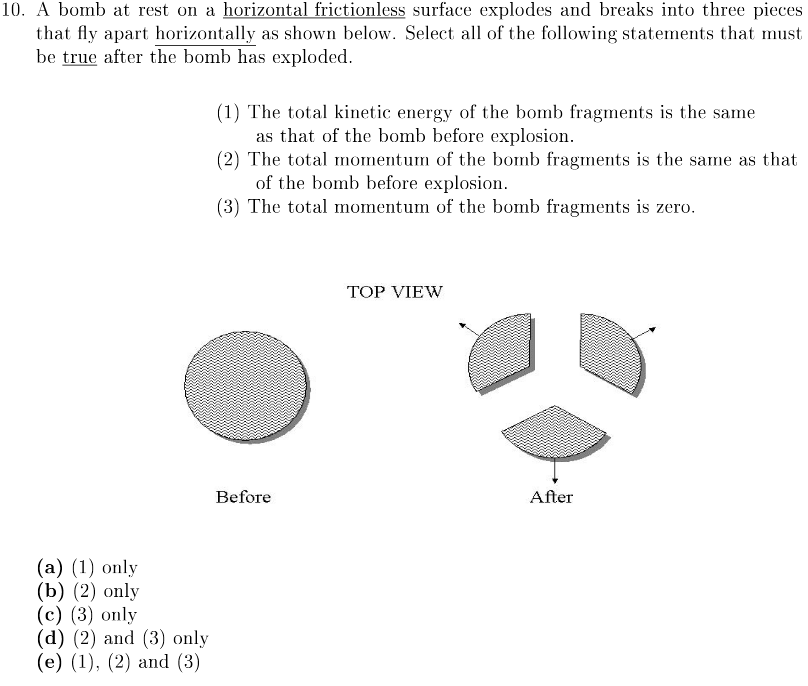}
    \caption{Q10 from the EMCS with the options provided for reference.}
    \label{fig:Q10}
\end{figure}

Q10 provides the diagram and options shown in Figure \ref{fig:Q10} and asks which statement(s) are true.
Here, 23\% of introductory physics students selected only option (2) (choice B). In other words, while these students recognized that the total momentum of the system is conserved, they did not realize that the initial momentum is zero, and thus the final momentum of the fragments is zero as well. Only 20\% of the TAs identified this difficulty, and 59\% of the TAs selected answer choice E, which stated that all the statements are correct. In other words, the TAs expected that students may think that in an explosion, the total kinetic energy is the same before and after the collision, but only 10\% of students selected this answer choice. One TA who selected choice E for this stated, ``Student[s] could forget that directions in KE [have] no meaning so the three pieces are moving in different directions; hence, their components cancel each other." This suggests that the TAs think that students have difficulty distinguishing between kinetic energy and momentum because they expect that students do not realize that kinetic energy is a scalar, and also, that they expect it to be conserved from before to after the explosion.

\subsubsection{Impulse-Momentum Theorem (Q18, Q19, Q23)}

Q18, Q19, and Q23 require use of the impulse-momentum theorem. On Q18 and Q19, the TAs performed poorly (PCK scores of 51\% and 49\%, respectively). On Q23, TAs exhibited moderate performance, receiving a PCK score of 62\%, as shown in Figure \ref{fig:EMCS_PCK_Responses}.

Q18 asks which statement is true about linear momentum. There were two alternate conceptions on this question (both moderate): 21\% of students thought that momentum is a scalar quantity (choice C) and 19\% of students thought that the SI unit of momentum is $kg\ m^{2}/s$ (choice D). TAs were only able to identify option D as being common (39\% of the TAs) with only 6\% of TAs selecting choice C. Furthermore, there are two options, choice A, which states that momentum is a force, and choice B, which states that the momentum of an object is always positive, which were selected by 21\% and 30\% of TAs, but only by 5\% and 10\% of students, respectively. This indicates that there is a major disconnect between most of the students and a majority of the TAs, since over half the TAs selected options that only a combined 15\% of students selected, and the most common alternate conception of students, namely that momentum is a scalar, is only identified by 6\% of the TAs.

Q19 asks why it would be better to hit a haystack than a concrete wall when the brakes of your bicycle fail. There was a moderate difficulty on this question: 25\% of students selected option A, which states that the haystack gives you a smaller impulse than the concrete wall. In other words, these students did not recognize that the impulse is the same in both cases because the change of momentum is the same. They likely did not think of applying the impulse-momentum theorem to realize that the haystack changes your momentum over a longer time (choice B). While 41\% of the TAs identified this difficulty, 30\% of the TAs selected answer choice D, which states that the change in momentum is smaller if you hit the haystack than if you hit the concrete wall, but only 9\% of students selected this option. One TA who selected choice D stated, ``People know that the haystack is the right answer and think that smaller momentum change will be less painful than a larger momentum change, but the bicycle will have the same momentum change regardless of the hitting the haystack or wall." This implies that TAs who selected choice D seem to think students who select choice D may think similarly to those who selected A, but may not recognize that the momentum change is the same. Furthermore, choices C and E were selected at rates of 4\% each by students, but were selected by 13\% and 11\% of the TAs, respectively. Overall, choices C, D, or E were selected by a combined 54\% of TAs, but only accounted for a combined 17\% of student difficulties. The students who struggled on this question tended to select choice A, but most of the TAs were not able to identify this answer choice as being common, showing a large disconnect between many students who struggled and a majority of the TAs.

In Q23, two balls of equal mass, but different material (one rubber, the other putty), are dropped from the same height above a horizontal surface. The rubber ball bounces after striking the surface, and the putty ball comes to rest. The question states that the velocity change from initial to final takes the same time and asks to compare the average forces on each ball.
There are two moderate difficulties on this question; 29\% of students selected that the average forces are the same (choice A) and 26\% of students selected that the average force on the putty ball is greater (choice C). TAs exhibited moderate performance on this question (62\% of the maximum possible), with 41\% of them selecting choice A and 23\% selecting choice C. One TA who selected choice A stated, ``Students would be aware of the relationship between impulse and the average force. However, they would ignore that the changes of the impulse are different for the rubber ball and the putty ball," implying that students might only consider the fact that the putty ball takes the same time as the rubber ball and not that the impulse is larger in the case of the rubber ball. One TA who selected choice C stated, ``They might think because the putty stuck...it has a greater impact," implying that students think that impulse has to do with the end result of the fall rather than how the forces change during the impact with the ground. 
However, a combined 25\% of the TAs selected answer choices D and E, which both imply that the question cannot be answered without more information (choice D --- to know the size of the balls; choice E --- need to know the height from which the balls are dropped). But only 6\% of students selected these answer choices. In other words, 25\% of TAs thought that students would focus on the lack of values when trying to answer this question rather than considering the physical principles applicable, but only a combined 6\% of students had this difficulty. This is similar to Q16, where 19\% of TAs thought that students would focus on lack of values, but only 2\% of students had this difficulty. This could be interpreted as another example in which a sizeable fraction of the TAs thought that students would be more novice-like than they actually are.

\subsubsection{Work-Kinetic Energy Theorem (Q8, Q24, Q25)}

Q8, Q24, and Q25 required using the work-kinetic energy theorem. On Q25, TAs performed poorly (PCK score 36\%); on Q8, they performed moderately (PCK score of 57\%) and on Q24, TAs performed well (PCK score of 90\%), as shown in Figure \ref{fig:EMCS_PCK_Responses}. We note that TAs' PCK score on Q25 was one of the lowest on the survey and their score on Q24 was the highest, suggesting that TAs' ability to identify alternate conceptions related to the work-kinetic energy theorem is context dependent, a finding that is consistent with 
%our 
previous papers on PCK using the FCI \cite{fciPCK} and the TUG-K \cite{tugkPCK}.

On Q8, a heavy block needs to be lifted through a height $h$ with a string and is pulled at constant velocity. Two options are given: lifting vertically or along a frictionless incline plane. The question provides five statements and asks which one is true. Option A states that the magnitude of the tension force is smaller when the block is lifted vertically compared to when it is lifted along a frictionless incline, and option B states that the magnitudes are the same. These two options combined were selected by only 14\% of students, but by 31\% of the TAs. In other words, roughly one third of the TAs think that students may not realize that the tension is smaller when the block is lifted along the frictionless incline than straight up. This suggests that they expect students to exhibit more novice-like thinking than they actually do. Instead, the most common alternate conception (29\% of students) is thinking that the work done is smaller when the block is lifted along the incline than when lifted vertically, which 51\% of the TAs were able to identify. These students likely realize that the tension is smaller when the incline is used and may confuse this with work (or not realize that the distance is longer for the incline, which compensates for the smaller force required to lift the box). One TA who selected this answer choice stated, ``The force is small in [the case of the incline] so by a misconception of work students may think work is smaller."

On Q24, two blocks are initially at rest on a frictionless horizontal surface. The blocks have different masses but are pulled by the same constant force. The question asks to compare the kinetic energies of the blocks after being pulled through the same distance. There are two moderate difficulties on this question; 26\% of students selected that the kinetic energy of the smaller block is greater because it achieves a larger speed (choice B), and 25\% of students selected that the kinetic energy of the larger block is greater because of its larger mass (choice C). TAs performed very well on this question, with 86\% of them selecting either of these two options. One TA who selected choice B stated, ``Since the applied force is the same and block A has a smaller mass, students might assume it will have a greater kinetic energy because it has a greater velocity." This implies that the students may recognize the force on the smaller block causes a greater acceleration and therefore a greater final velocity, causing the students to select choice B because they are only considering how fast a given block is going. One TA who selected choice C stated, ``[Students] may not realize they have different accelerations and hence different velocities." This implies that the students who selected choice C may only consider the mass difference and not the accelerations, thinking that the blocks would end up at the same speed and thus the kinetic energy of the larger block would be greater. It is noteworthy that, unlike other questions, very few TAs selected the responses that mention needing to know values (choices D and E --- selected by only 12\% of the TAs and 17\% of the students). This is in contrast to Q23 and Q16, where 25\% and 19\% of the TAs selected answer choices, which imply that more information is needed, but these answer choices were selected by only 6\% and 2\% of students, respectively.

On Q25, a box slides on a horizontal surface with friction and comes to a stop. The question asks to identify what is equal to the change in kinetic energy of the box. There is one moderate difficulty on this question: 20\% of students selected an option that states that the momentum of the box multiplied by the distance traveled before coming to rest is equal to the change in kinetic energy (choice A). The TAs did a poor job of identifying this difficulty, and only 29\% of TAs selected this choice. 
The incorrect answer choice selected most often by TAs is a choice which stated that the mass of the box multiplied by the deceleration of the box is equal to the change in kinetic energy (choice D). This was identified by 36\% of the TAs, however, only 12\% of students selected this incorrect answer choice.
One TA who selected choice D stated, ``I think that the most common incorrect answer would be [choice D] because [the students] are thinking about the force due to friction...but then they're missing out on the distance, which [is the connection between force and] kinetic energy change.'' This suggests that TAs who identified choice D expected students to confuse work with force, when very few students did so. The poor performance of TAs on this question is in large part due to the fact that the most commonly selected incorrect answer choice by TAs was the least commonly selected incorrect answer choice by students. 
The TAs' struggle was also evidenced by the fact that many TAs had difficulty coming up with ways in which students could get this question wrong, with 11 out of the 70 TAs either leaving the question blank, writing ``N/A,'' or indicating in their written responses that their choices were a guess because they felt that this question only required units to solve. For example, one TA who selected choice B said, 
``If they don't know dimensional analysis any choice could be possible."
This indicates that many TAs thought that students may not check the units and be able to recognize the answer immediately, thus causing them difficulty when attempting to determine the most commonly selected incorrect answer choice.

\subsubsection{Work Done by Gravity, (Q1, Q6)}

Q1 and Q6 required understanding the work done by gravity. On Q1, TAs performed poorly (PCK score of 39\%), and on Q6 they performed well (PCK score of 86\%), as shown in Figure \ref{fig:EMCS_PCK_Responses}.

In Q1, a suitcase is lifted from the floor to a table. The question asks to identify the factors (other than the weight of the suitcase) that determine the work done by the gravitational force and provides options: (1) whether it is lifted directly up or along a longer path, (2) whether it's lifted quickly or slowly, and (3) the height of the table above the floor. The most common difficulty here (34\% of introductory physics students) was to think that the path affects the work. This is somewhat similar to Q8, where students thought that the work done by the tension force on a block lifted along a frictionless incline would be smaller compared to if the block was lifted directly up (same height difference). On Q1, only 38\% of the TAs identified this alternate conception compared to 51\% on Q8, leading to their lower PCK performance on Q1 (39\%) compared to Q8 (57\%).
On Q1, 28\% of the TAs expected that students might not recognize that the height matters and instead think that it was just the path that affected the work (choice A), but only 6\% of students made this selection. One TA who selected choice A stated, ``Students will assume that the length of the path will affect the amount of work." This is another example where it could be interpreted that the TAs who selected answer choice A thought that students might exhibit more novice-like thinking than they actually do because these TAs did not realize that most students know that the height matters (94\% of students did).
Furthermore, 30\% of the TAs expected that students might think that the speed at which the suitcase is lifted affects the work done on it --- options D and E, which included statement (2). But these answer choices were selected by only 14\% of students. One TA who selected choice D said, ``Intuitively, it feels harder to lift quickly rather than slowly." Another TA who selected choice E said, ``Since we physically feel more tired if we lift the suitcase over a longer path or [lift the suitcase] faster, it is common to misunderstand that it means more work is done." This implies that TAs who selected either choice D or E thought many students would think about the physical action of moving the suitcase and not consider the physics principles relevant --- once again, an example of some TAs thinking that students exhibit more novice-like thinking than they actually do.

In Q6 a satellite moves around the Earth in a circular orbit at a constant speed, with the only force acting on it being Earth's gravitational force. The question asks to identify a true statement about the satellite moving between two points in the orbit among the following: (A) the gravitational potential energy changes, (B) the work done by the gravitational force is negative, (C) the work done by the gravitational force is zero, (D) the velocity of the satellite remains unchanged, and (E) none of the above. There is one major alternate conception on this question: 46\% of students selected D. TAs performed quite well (PCK score of 86\%) because the vast majority of them identified this alternate conception (84\% of the TAs selected D). One TA who selected D stated, ``The speed of the satellite remains constant, and it is tempting to confuse velocity with speed." This implies that most TAs were able to successfully identify that students recognize that velocity and speed are related, but then do not recognize that the velocity has direction and therefore will be changing as the satellite moves along the path. We should also note that this is the question on the EMCS with the highest percentage of students selecting a given incorrect answer, which makes the fact that so many TAs successfully identified it very encouraging. (Even more encouraging is that on the four questions with the highest percentages of students selecting an incorrect answer choice --- Q6, Q12, Q15, and Q22 --- TAs' PCK performance is good, suggesting that they did well in identifying student difficulties when 38\% of students or more gravitated toward one incorrect answer choice.)

\subsubsection{Mechanical Energy is Constant, (Q13, Q15, Q17, Q22)}

Q13, Q15, Q17, and Q22 require understanding situations in which total mechanical energy is constant and applying it correctly. On Q13, TAs performed poorly (PCK score of 52\%), whereas on Q15, Q17, and Q22, they performed well (PCK scores of 79\%, 77\%, and 75\%, respectively), as shown in Figure \ref{fig:EMCS_PCK_Responses}.

\begin{figure}[h]
    \centering
    \includegraphics[scale=0.75]{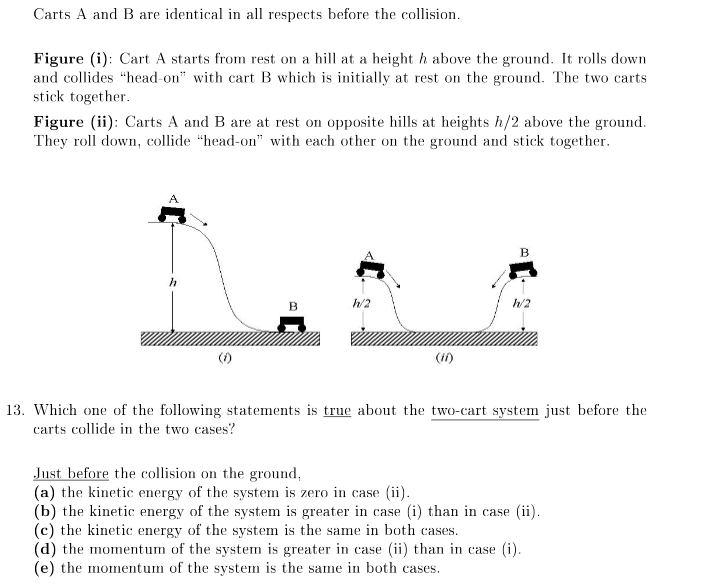}
    \caption{Q13 from the EMCS with the options provided for reference.}
    \label{fig:Q13}
\end{figure}

Q13 provides a diagram of two two-cart systems as shown in Figure \ref{fig:Q13}. The question asks to identify the statement that is true about the system just before the carts collide among (A) the kinetic energy of the system is zero in case (ii), (B) the kinetic energy of the system is greater in case (i) than in case (ii), (C) the kinetic energy of the system is the same in both cases, (D) the momentum of the system is greater in case (ii) than in case (i), and (E) the momentum of the system is the same in both cases. Here, 19\% of students either do not recognize that total mechanical energy is constant or misapply this and select choice B, with 39\% of the TAs identifying this alternate conception as common. 27\% of TAs selected answer choice E, thinking that students may not realize that the momentum is zero in case (ii) and thus it cannot be the same in the two cases. For example, one TA stated, ``Students may ignore the fact that the 2 carts in case (ii) move in the opposite direction, so they won't be aware that the momentum is zero in case (ii)." This implies that TAs who selected choice E think that students will not consider the vector nature of the momentum and instead think that the momenta of both carts in case (ii) will add to be the same as the cart A in case (i). However, only 12\% of students selected this choice.
Similarly, 17\% of TAs selected choice A, with one TA stating, ``[Students] may treat kinetic energy as vectors." This implies that these TAs thought that students who pick choice A would make the opposite mistake of those who picked choice E. Meaning these TAs thought students would think that kinetic energy acts as a vector when it is a scalar quantity, but only 8\% of students selected this choice.
Both choices A and E, being selected by a combined 44\% of TAs, imply that many TAs thought that many students would exhibit more novice-like thinking (kinetic energy is a vector, momentum is a scalar) than they do.

In Q15 two people of different masses are sliding down a frictionless slide with both starting from rest at the same height. The question asks to identify which choice best describes which person has a larger speed at the bottom of the slide. Here, 38\% of students thought that the heavier person would be faster at the bottom because their greater weight causes a greater downward acceleration (choice C). TAs performed well in identifying this difficulty (79\% of TAs selected option C, leading to a good PCK score of 79\%). One TA, for example, motivated their choice by stating that it was due to the ``misconception that heavier objects experience higher acceleration." This is encouraging, as it implies that the majority of TAs recognize that after instruction students retain the alternate conception that heavier objects have larger accelerations when sliding on ramps.

In Q17, a ball is dropped from a high tower and falls freely under the influence of the gravitational force. Students are then asked to identify the statement that is true about the situation among (A) the kinetic energy of the ball increases by equal amounts in equal times, (B) the kinetic energy of the ball increases by equal amounts over equal distances, (C) there is zero work done by the gravitational force as it falls, (D) the work done on the ball by the gravitational force is negative as it falls, and (E) the total mechanical energy of the ball decreases as it falls. There are two moderate difficulties on this question: 25\% of students select choice A and 27\% of students select choice D. The vast majority of the TAs identify these difficulties (66\% selected A and 16\% selected D), leading to a good PCK performance (77\%). One TA who selected choice A said, ``Speed increases by equal amounts in equal times, kinetic energy is related to speed." Another TA stated, ``The velocity increases with time linearly, but not for kinetic energy." This implies that TAs who selected choice A thought that students may use kinematics to recognize that velocity increases by equal amounts in equal times and select choice A as a result. One TA who selected choice D stated, ``Since force is in the negative direction, students might [overlook] the negative sign coming from [the displacement] and assume the work is negative." Similarly, another TA stated, ``Direction of gravitational force may not be taken into account." This implies that students who selected choice D may recognize that the ball must be falling downwards and associate that with negative work, rather than recognizing the direction of either the force of gravity or the displacement.

\begin{figure}[h]
    \centering
    \includegraphics[scale=0.75]{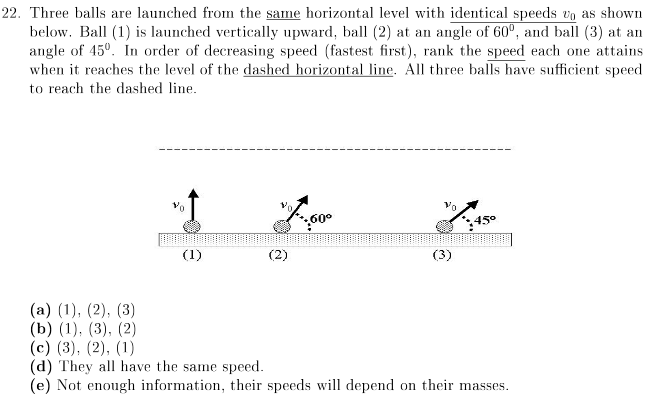}
    \caption{Q22 from the EMCS with the options provided for reference.}
    \label{fig:Q22}
\end{figure}

Q22 provides the diagram shown in Figure \ref{fig:Q22} and states that all three balls are launched from the same horizontal level with identical speeds. The question asks to rank the speed each ball attains when it reaches the dashed horizontal line. There is one major difficulty and one moderate difficulty on this question: 44\% of students selected option A --- (1), (2), (3) and 20\% of students selected options C --- (3), (2), (1). Here, 70\% of the TAs identified the major difficulty. For example, one TA stated, ``[Students] may be looking at the vertical velocity component." This implies that these TAs are able to recognize that students may focus on the fact that the vertical velocity component of each ball would be ranked in the order given, then select choice A as a result. A few of the TAs identified the moderate difficulty --- 13\% of the TAs selected choice C. One TA who did so stated, ``People think the sideways speed doesn’t change, but the vertical one is being slowed down more." Similarly, another TA stated, ``[Students will] consider the horizontal speed." This implies that these  TAs thought students might focus on the horizontal component of the velocity staying the same and would select the ranking that is based on this component, namely (3), (2), (1). We note that similar to other questions, Q22 provides an answer choice that the mass is needed in order to figure out the ranking, i.e., not enough information (choice E). But here, unlike in some other questions discussed earlier, very few TAs selected this answer choice (9\%), and it's reassuring that many TAs are able to successfully identify either the major or moderate student difficulty on this question, particularly because this is the question with the lowest student performance (only 28\% of students answer it correctly).

\subsubsection{Work Done by a Non-Conservative Force (Q9, Q12)}

Q9 and Q12 require understanding work done by non-conservative forces. TAs' PCK performance was good on both questions (69\% and 74\%), as shown in Figure \ref{fig:EMCS_PCK_Responses}.

On Q9 three bicycles approach a hill in three separate situations: (1) Cyclist 1 stops pedaling at the bottom, and her bicycle coasts up the hill, (2) cyclist 2 pedals so that her bicycle goes up the hill at a constant speed, (3) cyclist 3 pedals harder so that her bicycle accelerates up the hill. The question states that the retarding effects of friction can be ignored and then asks to identify all the cases in which the total mechanical energy of the cyclist and bicycle is constant. There is one moderate difficulty on this question: 21\% of students selected an option that states that case (2) is the only one in which mechanical energy is constant (choice B), likely confusing mechanical energy with kinetic energy. TAs did an okay job of identifying this difficulty: 37\% selected choice B; one TA who did stated, ``[Choice B] contains the keyword constant speed, therefore I believe this is a good distractor for students who are confused on conservation of energy." Another TA who selected choice B stated, ``It sounds like kinetic energy is conserved, but doesn't account for potential energy change." This implies that these TAs recognize that students might see constant speed and associate that with constant total mechanical energy, without recognizing that the potential energy is increasing as the cyclist moves up the hill. Many TAs were also able to identify other less prevalent difficulties: 27\% of TAs selected choice C, which states that cases (1) and (2) both have constant total mechanical energy (selected by 14\% of students) and 29\% of TAs selected choice E, which states that all cases have constant total mechanical energy (selected by 14\% of students). These suggest that as a group, the TAs are able to recognize that there are multiple types of incorrect reasoning students can use on this question and were able to account for them accordingly.

In Q12 a box is pulled along a horizontal surface, with friction $F_k$, via a constant horizontal force $F_A$ resulting in constant speed. The question asks to identify what statement is true about the motion of the box. There is one major alternate conception on this question: 40\% of students think that $F_A>F_k$ and 64\% of the TAs identified this. One TA who did stated, ``Some students might think that the box is moving forwards, so $F_A$ is greater than $F_k$ (might not put much emphasis on constant speed corresponding to constant kinetic energy)" and another TA stated, ``Since the box moves it may be tempting to say that the force applied is greater [in order to create the movement], but [the forces can be equal], and the box can move at constant velocity." This implies that most of the TAs recognize that many students have the alternate conception that an object moving at a constant velocity must have a nonzero net force acting on it.

\subsection{Overall Findings for TA PCK vis a vis RQ1}

Awareness of students' common difficulties and being able to understand student reasoning about why they are having those difficulties after instruction are important aspects of pedagogical content knowledge. An instructor can take advantage of this knowledge of students' common difficulties and use it to assist with the development of pedagogical approaches that assist in student learning \cite{tugkPCK, fciPCK, thompson2011preparing, sadler2013middleschool}.
Our investigation used the EMCS instrument to evaluate this aspect of pedagogical content knowledge relating to energy and momentum concepts for 70 TAs, who were all first-year physics graduate students enrolled in a TA professional development course. For each item on the EMCS, the TAs were asked to identify what incorrect choice they expected to be the most commonly selected by introductory physics students after traditional instruction and  provide a short written explanation to justify why they selected the choice they did. Additionally, think-aloud interviews were conducted to obtain more in-depth accounts of reasoning TAs use to decide what alternate conceptions are most prevalent for each item on the survey. 

Below, we provide a brief overview of how our results aligned with our research goals, organized by performance and concept.

\vspace{0.2 in}

\subsubsection{\bf{RQ1-A: Concepts in which TAs performed poorly when identifying alternate conceptions}}
\vspace{0.2 in}

For three of the seven concept groups involved in the EMCS, TAs performed poorly (received a normalized average PCK score of less than 52\%) when identifying alternate conceptions based on their normalized average PCK scores. 

{\bf{Conservation of Momentum:}}
In both questions discussed relating to conservation of momentum, many TAs were not able to identify students' alternate conceptions. Instead, many expected certain incorrect answers to be common when they were not. For example, Q10, in which TAs overwhelmingly believed the students may not distinguish between an elastic and an inelastic collision. For both of the questions, the TAs' responses indicated that many of them thought the students would exhibit more novice-like thinking than they did. This indicates a disconnect between students and TAs on what the students struggle with relating to the conservation of momentum.

{\bf{Mechanical Energy is Constant and Momentum is Conserved:}}
In both questions discussed relating to total mechanical energy being constant and momentum being conserved for the system under consideration, we see examples of common alternate conceptions held by students that many TAs were not able to identify. Instead, TAs' selection of the most common incorrect answer choice was spread over a few different options rather than converging overwhelmingly on one. For example, on Q3, comparable percentages of TAs selected the most common incorrect answer choice and another answer choice that was selected by few students, suggesting that many TAs did not have a clear understanding of students' common alternate conceptions. Similar to the questions on conservation of momentum, many TAs' responses indicate that they expected students to exhibit more novice-like thinking than they actually did. For example, on Q16, 19\% of the TAs expected that students may not consider the relevant physics principles applicable and gravitate toward the answer choice that implies that more information is needed, i.e., specific numbers, but only 2\% of students fit this category.

{\bf{Impulse-Momentum Theorem:}}
In two of the three questions related to the impulse-momentum theorem, there are alternate conceptions that were not common among students, but many TAs expected them to be common, leading to poor PCK performance. In the third question in this group (Q23), TAs' performance was moderate, while the majority of TAs identified one of the two alternate conceptions, 25\% of TAs who selected answer choices that only a combined 6\% of students selected, bringing down their average score greatly.

\vspace{0.2 in}
\subsubsection{\bf{RQ1-B: Concepts in which TAs did well when identifying alternate conceptions}}
\vspace{0.2 in}

For two of the seven concept groups involved in the EMCS, TAs performed well (received a normalized average PCK score of greater than 67\%) when identifying alternate conceptions.

{\bf{Mechanical Energy is Constant:}}
Out of the four questions discussed relating to total mechanical energy remaining constant for a given system, we see one example of a question in which there is an alternate conception that was not common among students, which many TAs expected to be common (Q13), and in the other three questions, there are examples of alternate conceptions that were common among students, which the majority of the TAs were able to identify (79\%, 66\%, and 70\% of the TAs were able to identify the most common alternate conceptions in Q15, Q17, and Q22, respectively). It is encouraging that in most of the questions relating to total mechanical energy being constant, many of the TAs were able to identify students' alternate conceptions, suggesting that they had good intuition about where students struggle with this concept.

{\bf{Work Done by a Non-Conservative Force:}}
In both questions related to the work done by a non-conservative force, most of the TAs were able to identify students' alternate conceptions. For example, on Q12, 64\% of the TAs identified that students would think that motion at constant velocity implies a nonzero net force in the direction of motion.

\vspace{0.2 in}

\subsubsection{\bf{RQ1-C: Concepts in which TA performance was context dependent}}

\vspace{0.2 in}

For two of the seven concept groups involved in the EMCS, TAs were able to identify the common student alternate conceptions well in some questions but struggled in other questions.

{\bf{Work-Kinetic Energy Theorem:}}
This group of three questions proved quite interesting in that TA performance was extremely varied: on Q25, they had one of the lowest PCK performance of the survey, on Q8, their performance was moderate; and on Q24, they had the highest PCK performance of all questions. 
This is because on Q25, the majority of the TAs were not able to identify the most common alternate conception. On Q8, only around half of the TAs identified the most common alternate conception. Finally, on Q24, the vast majority of the TAs identified one of the two alternate conceptions. While previous studies have found context dependence of TA PCK performance in prior studies \cite{fciPCK,csemPCK, tugkPCK}, such extreme variation has typically not been observed before.

{\bf {Work Done by Gravity:}} These two questions provide another example of context dependence of TA performance: Q1 --- poor performance, Q6 --- high performance (second highest on the entire EMCS). On Q1, only 38\% of the TAs identified the most common alternate conception of 34\% of students, and the rest of the TAs selected answer choices that combined for only 20\% of student responses. However, on Q6, 84\% of the TAs identified students' alternate conception, though this may be due to the high prevalence of this alternate conception (46\% of students).

\subsection{RQ2: Comparison with Previous Studies which Use Similar Methods}

First, looking broadly at TAs' PCK performance on the EMCS, we find that out of 18 questions that have common student alternate conceptions, the TAs performed poorly on nine questions, moderately on two questions, and well on seven questions. This is in contrast to the other two studies in which this type of analysis has been conducted: in the context of the FCI \cite{fciPCK}, out of the 23 questions with common student difficulties, TA performance was poor on 6 questions, moderate on 6 questions, and good on 11 questions; in the context of the TUG-K \cite{tugkPCK}, out of the 17 questions with common student difficulties, TA performance was poor on 6 questions, moderate on 5 questions, and good on 6 questions. Based on this, it appears that TA PCK performance on the EMCS was generally worse (poor performance on 50\% of the questions) than their performance on the FCI (poor performance on 26\% of the questions) and TUG-K (poor performance in 35\% of the questions). (We note that if we use the cutoff of 52\% for poor PCK performance in the other two studies, the conclusion is the same; the percentage changes only for the FCI from 26\% of the questions having poor performance to 30\%).

One of the main findings when analyzing TAs' approaches to identifying common student alternate conceptions in the EMCS is that they often expected students to exhibit more novice-like thinking than they actually did. Looking back at the previous studies where we used a similar methodology \cite{fciPCK,tugkPCK,csemPCK}, we find similar results for the FCI \cite{fciPCK}. In particular, on the FCI, there were six questions in which TAs' PCK performance was poor (Q13, Q15, Q16, Q22, Q25, Q29), and in five of these questions (Q13, Q15, Q16, Q25, Q29), TAs' poor performance can be partially explained as the TAs expecting that students would exhibit more novice-like thinking that they did. For example, on Q13, a boy throws a steel ball straight up, and the question asks to identify all the forces acting on the ball after it leaves the boy's hand and before it touches the ground. The most common alternate conception (50\% of students) was related to the impetus view of motion: a steadily decreasing upward force acts on the ball as it is going up, and both on the way up and on the way down, there is a constant force of gravity. Only 16\% of the TAs selected this answer choice. Instead, 44\% of the TAs gravitated toward an answer choice that included the steadily decreasing upward force on the way up, but on the way down, the force of gravity steadily increases as the ball gets closer to the Earth (11\% of students). These TAs were aware that students would think that there is a decreasing force on the way up (impetus view), but they expected that students may also think that the force of gravity increases as objects get closer to the Earth. In other words, they expected students to exhibit more novice-like thinking than they actually did.

Also, on the FCI, there were six questions (Q5, Q11, Q12, Q21, Q23, Q28) in which TAs' performance was moderate, and in four of these questions (Q5, Q11, Q12, Q28), the TAs' performance can be partly explained by the same reason. For example, on Q11 a puck is sliding along a horizontal frictionless path, and the question asks for the forces acting on it. On this question, 40\% of the TAs selected answer choices that did not include the normal force, but these answer choices were only selected by 13\% of students. In other words, many TAs thought that students may not recognize that the surface exerts a force on the puck, which was not the case.

In short, on the FCI, in 9 out of the 12 questions on which TAs' performance was lower than 67\% (i.e., poor or moderate performance), a close look at TAs' choices suggests that they thought students were more novice-like than they actually were.

\begin{figure}
    \centering
    \includegraphics[width=0.5\linewidth]{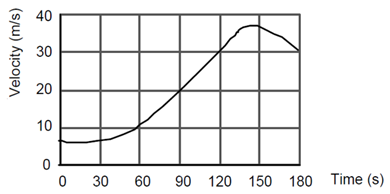}
    \caption{Figure provided for Q6 on the TUG-K.}
    \label{fig:TUG-K6}
\end{figure}

For the TUG-K \cite{tugkPCK}, findings are not as clear-cut, in part because the survey is related to graphing and interpreting graphs, and it is sometimes difficult to interpret certain answer choices as more novice-like than others, though it is possible in some cases. For example, Q6 provides the velocity vs time graph shown in Figure \ref{fig:TUG-K6} and asks to calculate the acceleration of the object at time $t=90s$. Here, one of the answer choices was $9.8 m/s^{2}$ and 20\% of the TAs selected this answer choice, whereas only 6\% of students selected it. In other words, 20\% of the TAs expected that students might not attempt to do any calculation based on the graph and simply select the acceleration due to gravity, but almost no students did. (The most common incorrect approach was to divide $20 m/s$ by $90 s$, i.e., attempt to use an equation like $a=v/t$ --- 40\% of students did this, but only 20\% of the TAs identified it as the most common incorrect approach.)

Q7 is very similar: it also provides a velocity vs time graph and asks for the acceleration at a specific time, and it too provides an answer choice of $9.8 m/s^{2}$. Similar to Q6, on Q7, 28\% of the TAs expected that this answer choice would be the most common, but only 6\% of students selected it.

The CSEM is a more difficult survey for students \cite{csemPCK}, and the TAs also seemed to have less intuition about what kind of reasoning students could use to arrive at incorrect answers. 
The CSEM study \cite{csemPCK} includes a comparison with the FCI and TUG-K studies which points out that TAs found the PCK task to be more challenging in the context of the CSEM compared to the FCI or TUG-K. It is stated that this could be because ``our daily experience with the real world leads to a relatively predictable (Aristotelian) world view, and TAs could more easily reason their way to common misconceptions held by students. Electricity and magnetism, on the other hand, deals with concepts that are not primarily learned experientially (e.g., charges, fields, and currents), which likely makes it more difficult to predict the [common alternate conceptions] of students." 
This difficulty can lead to the TAs selecting answer choices that are due to more novice-like reasoning than students may have, and there are a few examples of that (more below). However, it is also noted that \cite{csemPCK} when engaged in the CSEM PCK task, the TAs often ``selected answer choices which incorporate both correct and incorrect ideas" \cite{csemPCK}, and this can lead to the TAs expecting that students are less novice-like than they actually are (more below). But also, due to the CSEM being more difficult for students, there are more questions with multiple distractors compared to the FCI and TUG-K, making it more difficult to identify specific instances in which the TAs expect students to be more or less novice-like than they actually are.

\begin{figure}
    \centering
    \includegraphics[width=0.5\linewidth]{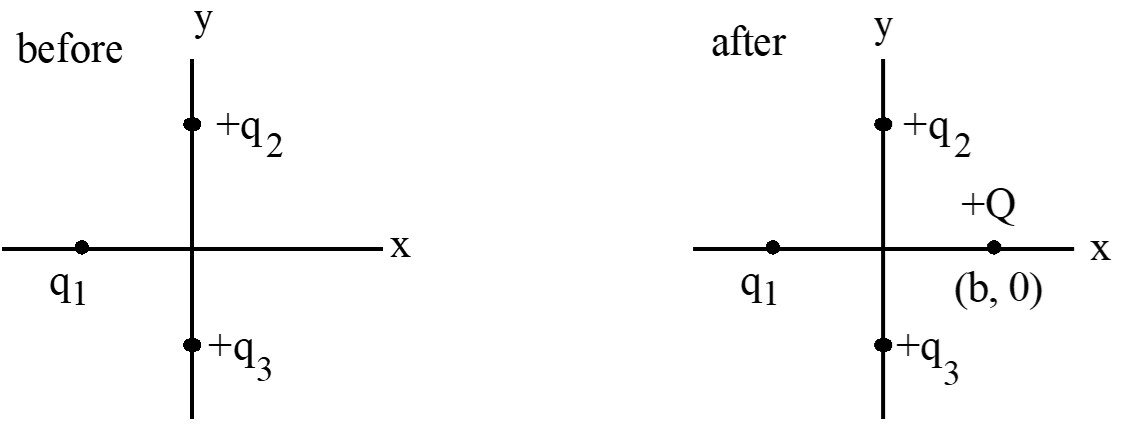}
    \caption{Figure provided for Q8 on the CSEM.}
    \label{fig:CSEM8}
\end{figure}

However, there are still questions that are in both of these categories. For example, questions 8 and 9 on the CSEM are analogous to some of the EMCS questions in that they provide answer choices that state that more information is needed when the questions can be answered with the information provided. CSEM question Q8 provides the figure shown in Figure \ref{fig:CSEM8} and states that in the figure on the left, the charges (all positive) are such that the net electrostatic force on $q_1$ points to the left and asks what will happen to the net force on $q_1$ if a new positive charge $+Q$ is placed at the position shown in the figure on the right. One of the answer choices states that the answer cannot be determined without knowing the magnitude of $q_1$ and/or $Q$. Here, 27\% of the TAs expect this answer choice to be the most common, but only 8\% of students select it. Q9 is very similar, and 23\% of the TAs select the answer choice that states more information is needed, but only 5\% of students selected it.

Q21 is another example: this CSEM question asks what happens to a positive charge placed at rest in a uniform magnetic field. One of the answer choices corresponds to confusing a magnetic with an electric field and states that the charge will move with a constant acceleration since the force has a constant magnitude. This answer choice was selected by 31\% of the TAs but only 8\% of the students. Instead, students were more likely to think that the charge will move around a circle at constant speed (21\% of the students, 41\% of TAs). In other words, 31\% of the TAs expected that students are more novice-like than they actually are.

And yet on other CSEM questions, the TAs expected students to be less novice-like than they actually are. For example, on CSEM question Q2, a small amount of negative charge is placed at some point P on a neutral hollow sphere made of electrically insulating material. Here, 21\% of the students do not distinguish between insulating and conducting and think that the charge has distributed itself evenly over the outside of the sphere, and 25\% of the TAs identify this difficulty as being the most common. However, 49\% of the TAs select the answer choice that states that most of the charge is at point P, but some spreads over the sphere, which is what would happen if the sphere was not perfectly insulating, i.e., it's an answer that is ``closer" to the correct answer than confusing insulating with conducting. But only 5\% of students select this answer choice.

There are a few other examples, but in general, for the CSEM, it was more difficult to find instances in which the TAs expected students to be either more or less novice-like than they were for the aforementioned reasons.

\section{Summary and Key Findings}

Our findings reveal that even motivated physics graduate TAs struggle to understand how introductory physics students reason about physics concepts. This gap between physics TA assumptions and the reality of student difficulties can have profound implications for helping students develop a robust knowledge structure. TAs who overestimate student difficulties may design activities that are insufficiently challenging, failing to promote the cognitive engagement necessary for deep learning. Conversely, those who underestimate challenges may create activities that frustrate and discourage introductory physics students. 
The solution may lie in developing TA capacity to understand and respond to student thinking. The approach to professional development discussed in this research addresses TA understanding of student difficulties by making student thinking visible through analysis of conceptual assessment data related to energy and momentum. This can challenge TAs' incorrect assumptions about student difficulties by comparing predictions with reality and building pedagogical reasoning through interpretation of patterns in introductory student thinking, e.g., via in-class discussions.

For example, often on questions on which the TAs performed poorly, they expressed that they thought students would make mistakes that were more novice-like than what the students actually did. This can be seen in examples like Q5, Q8, and Q10 in which TAs' performance was below 60\%. This implies that many TAs tend to think that students will not have some of the baseline knowledge, like understanding the difference between types of collisions or, for example, the difference between the magnitude of the tension force when an object is pulled directly upwards or along an incline plane.  
Although these topics are covered in introductory physics courses relating to energy and momentum, many TAs still expected that introductory physics students would retain novice-like alternate conceptions after instruction, as if they had not learned anything about a given topic.

Encouragingly, on the majority of the questions in which there was a major student difficulty, many TAs were able to identify it: overall, they performed well on 4 out of the 6 questions on which there's a major student difficulty. In contrast, the TAs performed well on only 3 out of the 12 questions which have at most a moderate difficulty. This implies that many TAs are more able to recognize introductory physics student alternate conceptions when they are very common. This is in contrast with the TUG-K \cite{tugkPCK}, in which many TAs struggled more on the questions that had major difficulties, suggesting that some TAs have a more difficult time identifying major difficulties in the context of kinematics graphs than in the context of energy and momentum. This suggests that the findings from one study on TAs' ability to identify student alternate conceptions are hardly generalizable, and research should be conducted with other surveys to examine the extent to which TAs are knowledgeable about introductory physics student alternate conceptions across different physics topics.

With regard to professional development programs, there are a few examples discussed in the introduction \cite{wittmann2002,Lampley2018PCKBio,park2025PCKmath,MariesTAs2020, thompson2011preparing}, which incorporate activities designed to help TAs learn about common introductory physics student difficulties. Given the prominence of PCK in the research literature on teachers at the K-12 level and the fact that one of the important parts of PCK is knowledge of student difficulties, such activities are surprisingly rare from descriptions of many professional development programs, specifically for graduate TAs who are an essential part of college education in the United States. We believe the PCK activity presented here, along with the data for both graduate TAs and students, can be particularly beneficial for developing TAs' knowledge of common student difficulties. Two of the authors have extensive experience with using PCK tasks in professional development programs. These kinds of tasks are especially beneficial for TAs, particularly if they are followed by sharing data from introductory students and discussing common student alternate conceptions. Additionally, data discussed here that includes information about what introductory student alternate conceptions TAs struggle to identify are also useful in determining which questions on a conceptual survey can be particularly conducive to productive TA discussions about student difficulties. Q3 is a good example because there is only one common alternate conception, but the TAs are split between three answer choices. Thus, in a group discussion, it is likely that TAs would come up with reasons why students may select multiple answer choices, and arguing for their position will help them develop a better grasp of student thinking. As noted in earlier studies \cite{fciPCK,tugkPCK}, TA discussions lead them to converge on a more common incorrect answer choice and thus learn about student alternate conceptions in the process.

\section{Conclusions and Future Directions}

By investing in effective professional development that helps TAs understand introductory physics student thinking as described here, physics departments can create cascading effects, i.e., better prepared TAs become more effective instructors as TAs, who may become future faculty members committed to evidence-based teaching. The path to effective physics education requires fundamental shifts in how we prepare educators. In particular, we should move beyond simply providing educators with technical training, e.g., in logistical details for how to teach them and help them develop a robust understanding of introductory student thinking, and the research discussed here in our previous articles using a similar methodology \cite{fciPCK,tugkPCK,csemPCK} can provide a good starting point. This is important in order to prepare educators who don't just transmit knowledge but facilitate effective learning experiences for students. 
In particular, these studies can provide useful activities to help TAs understand student reasoning. Future work may consider investigating other activities that can assist in developing TAs' understanding of student reasoning and investigate longitudinal effects of these activities on early educators.

This study showcases an activity which can assist with the development of educators' understanding of introductory physics student reasoning, which is a fundamental aspect of effective instructional design. The gap between what TAs assume about introductory physics student cognition and how students actually think represents both a challenge and an opportunity. By making this gap visible and providing structured support for developing PCK, we can prepare introductory physics educators who are capable of implementing active learning that promotes robust understanding using evidence, e.g., of student difficulties to continuously improve their teaching. The methodology presented here uses conceptual assessments to evaluate and develop PCK and provides a practical, scalable approach to effective teacher preparation. While demonstrated in physics, the principles and practices we described here can be applicable across STEM disciplines.

Future research should explore how the findings presented here can be adapted across different institutions in TA 
professional development programs. Although we hypothesize that TAs at large public research universities in the United States like the one presented here would have similar PCK about the EMCS concepts, any differences in findings across different types of institutions across the US and internationally would be useful in shedding light on how different the teaching assistants' PCK is about the EMCS concepts across different types of institutions. Similar studies that focus on the confidence and consistency of individual TAs may provide further insights about the abilities of TAs to identify these common difficulties. It may also be insightful to complete a follow-up study that identifies the same TAs' PCK after a set amount of time to see how consistent they are with their responses and/or to check retention of common student difficulties if they were specifically told them. It may be interesting to see how much TAs retain their knowledge about these difficulties a few weeks after the activity is completed or even after the professional development course is over. This may provide insight relating to how often a task like this should be done in order to prepare TAs for a semester of teaching, we hypothesize that it may be relevant for them to repeat this exercise at least a few times as a reminder of common student difficulties before they retain a significant amount of the common difficulties which students may have. 

\section*{Acknowledgments}

We are very grateful to all participants who helped with this research as well as professor Robert P. Devaty for helpful feedback on the manuscript.

\bibliography{refs}

\end{document}